\documentclass[12pt]{article}
\usepackage{graphicx}
\usepackage{siunitx}
\usepackage{lineno}
\usepackage{xspace}
\usepackage{subfigure}
\usepackage{csquotes}
\usepackage{document-defs}
\usepackage{url}

\newcommand\pubnumber{}
\newcommand\pubdate{\today}

\def\institute{Bergische Universitaet Wuppertal, Germany}
\def\support{\footnote{E-mail: arwa.bannoura@cern.ch}}

\def\Title#1{\begin{center} {\Large #1 } \end{center}}
\def\Author#1{\begin{center}{ \sc #1} \end{center}}
\def\Address#1{\begin{center}{ \it #1} \end{center}}

\newcommand\pubblock{\rightline{\begin{tabular}{l} \pubnumber\\
         \pubdate  \end{tabular}}}
\newenvironment{Abstract}{\begin{quotation}  }{\end{quotation}}
\newenvironment{Presented}{\begin{quotation} \begin{center} 
             PRESENTED AT\end{center}\bigskip 
      \begin{center}\begin{large}}{\end{large}\end{center} \end{quotation}}

\begin{document}

\begin{titlepage}
\pubblock

\vfill
\Title{Measurement of the inclusive \ttbar cross-section in the lepton+jets
channel in $pp$ collisions at $\sqrt{s}$ = 8 TeV with the ATLAS detector}
\vfill
\Author{ Arwa Bannoura\support \\ On behalf of the ATLAS Collaboration}
\Address{\institute}
\vfill
\begin{Abstract}
The inclusive \ttbar production cross-section is measured in
the lepton+jets channel using \SI{20.2}{fb^{-1}} of proton--proton collision data at
a centre-of-mass energy of \SI{8}{\TeV} recorded with the ATLAS detector at the
LHC.
Major systematic uncertainties due to the modelling of the jet energy scale and
$b$-tagging efficiency are constrained by separating selected events into three
disjoint regions. In order to reduce systematic uncertainties in the most
important background, the \Wjets process is modelled using \Zjets events in
a data-based approach.
The inclusive \ttbar cross-section is measured with a precision of
\SI{5.7}{\percent} to be $\sigma_{\ttbar} = 248.3 \pm 0.7 \,
(\mathrm{stat.}) \pm 13.4 \, (\mathrm{syst.}) \pm 4.7 \, (\mathrm{lumi.})~\text{pb}$, assuming a
top-quark mass of \SI{172.5}{\GeV}. The result is in agreement with the Standard
Model prediction.
\end{Abstract}
\vfill
\begin{Presented}
$10^{th}$ International Workshop on Top Quark Physics\\
Braga, Portugal,  September 17--22, 2017
\end{Presented}
\vfill
\end{titlepage}
\def\thefootnote{\fnsymbol{footnote}}
\setcounter{footnote}{0}

\section{Introduction}
The top quark is an elementary particle in the Standard Model of particle
physics. Being the most massive known elementary particle, makes it very
important to study its properties. These studies provide both a precise probe of
the Standard Model and a window for physics beyond the
Standard Model. In proton-proton collisions, the dominant production
process of top quarks is pair production (\ttbar production) via the strong interaction. 
The measurement of the production cross-section reported here is performed in the semileptonic decay
mode (lepton+jets), where one $W$ boson decays leptonically and the other $W$
boson decays hadronically, i.e. ($\ttbar \ \rightarrow \ell \nu b +
q\bar{q}'\bar{b}$)~\cite{TOPQ-2016-08}. The analysis is based on
data collected with the ATLAS detector~\cite{PERF-2007-01} at the LHC
corresponding to an integrated luminosity of \SI{20.2}{fb^{-1}} at a $pp$ centre-of-mass energy of $\sqrt{s} = \SI{8}{\TeV}$.
The measurement is in agreement with the Standard Model theoretical prediction,
which is $\sigma(pp \rightarrow \ttbar) \ = \ 253^{+13}_{-15}\ \ \
\mathrm{pb}.$~\cite{Cacciari:2011hy}.

\section{Event selection}
\label{sec:sel}
\ttbar events in the lepton+jets channel are identified to include electrons
and muons, jets, some of which are $b$-tagged for likely containing $b$-hadrons,
and sizable missing transverse energy.
Selected events are required to have one isolated charged lepton, i.e.
an electron or muon, with transverse momentum $p_T > \SI{25}{\GeV}$. Electrons
are required to be within pseudo-rapidity $|\eta| < \num{2.47}$, excluding the calorimeter overlap-region
$\num{1.37} < |\eta| < \num{1.52}$. Muons are selected within $|\eta| <
\num{2.5}$. Events must have at least four jets with $p_T > \SI{25}{\GeV}$ and $|\eta| <
\num{2.5}$. At least one of the jets has to be $b$-tagged. To enhance the fraction of events
with a leptonically decaying $W$ boson, events are required to have \MET $>
\SI{25}{\GeV}$ and the transverse mass \mtw of the lepton - \met system is
required to be \mtw $> \SI{30}{\GeV}$.
\section{Backgrounds estimation and modelling}
The dominant background to \ttbar production is \Wjets production.
This analysis uses a sample defined from collision data to model the discriminant
distribution shapes for this background, while the normalisation is determined
in the final fit. The multijet background is also modelled using
collision data but normalised using control regions. All other backgrounds, i.e.
single top, \Zjets and diboson production, are determined using simulated events
and theoretical predictions.

The method to obtain a modelling of the \Wjets background shape from data is
based on the similarity of the production and decay of the $Z$ boson to that of
the $W$ boson. An almost background-free \Zjets sample is selected from data by
requiring events to contain two opposite charged leptons of the same flavour,
i.e.
$e^+ e^-$ or $\mu^+ \mu^-$. The dilepton invariant mass $m(\ell\ell)$ has to
match the $Z$-boson mass ($80 \leq m(\ell\ell) \leq \SI{102}{\GeV}$). These
events are then \enquote{converted} into \Wjets events.
This is achieved by boosting the leptons of the $Z$-boson decay
into the $Z$ boson rest-frame, scaling their momenta to that of a lepton decay
from a $W$ boson by the ratio of the boson masses ($m_W/m_Z$) and
boosting the leptons back into the laboratory system. After this conversion, one
of the leptons is randomly chosen to be removed, and the \metvec is re-calculated.
Finally, the event selection cuts discussed in Sect.~\ref{sec:sel} are applied,
except for the $b$-tagging requirement.
In the following, this sample is referred to as the \ZtoW sample. 

In order to construct a sample of multijet background events, 
different methods are adopted for the electron and muon channels. The
\enquote{jet-lepton} method~\cite{ATLAS-CONF-2014-058} is used in the electron
channel to model the background due to fake electrons using a dijet sample,
where a jet that passes the selection cuts of a signal electron is selected to
resemble the electron. The \enquote{anti-muon} method~\cite{ATLAS-CONF-2014-058}
is used in the muon channel. A dedicated selection on data is performed to
enrich a sample in events that contain fake muons, by changing some of the muon
identification cuts. The normalisation of the multijet background is obtained
from a fit to the observed \MET in the electron channel or \mtw~distribution in
the muon channel.
\section{Systematic uncertainties}
Several sources of systematic uncertainties affect the \ttbar cross-section measurement. 
In addition to the luminosity determination, they are related to the modelling
of the physics objects, the modelling of \ttbar production and the understanding of the
background processes. All of them affect the yields and kinematic
distributions (shape of the distributions) in the three signal regions. The
systematic uncertainties are evaluated using pseudo-experiments. Details
about the sources and evaluation of systematic uncertainties are explained
in~\cite{TOPQ-2016-08}.
\section{Extraction of the \ensuremath{\ttbar} cross-section}
Selected events are separated into three disjoint signal regions. \SRone ($\ge4$
jets, 1 $b$-tag) has the highest background fraction and the highest selected
signal events. \SRtwo (4 jets, 2 $b$-tags) provides an unambiguous
association of the reconstructed objects to the top-quark decay products. \SRthr
($\ge4$ jets, $\ge 2$ $b$-tags excluding events from \SRtwo) has the smallest
background fraction.

\begin{figure}[!htbp]
	\centering
	\subfigure[]{
	\includegraphics[width=0.45\textwidth]{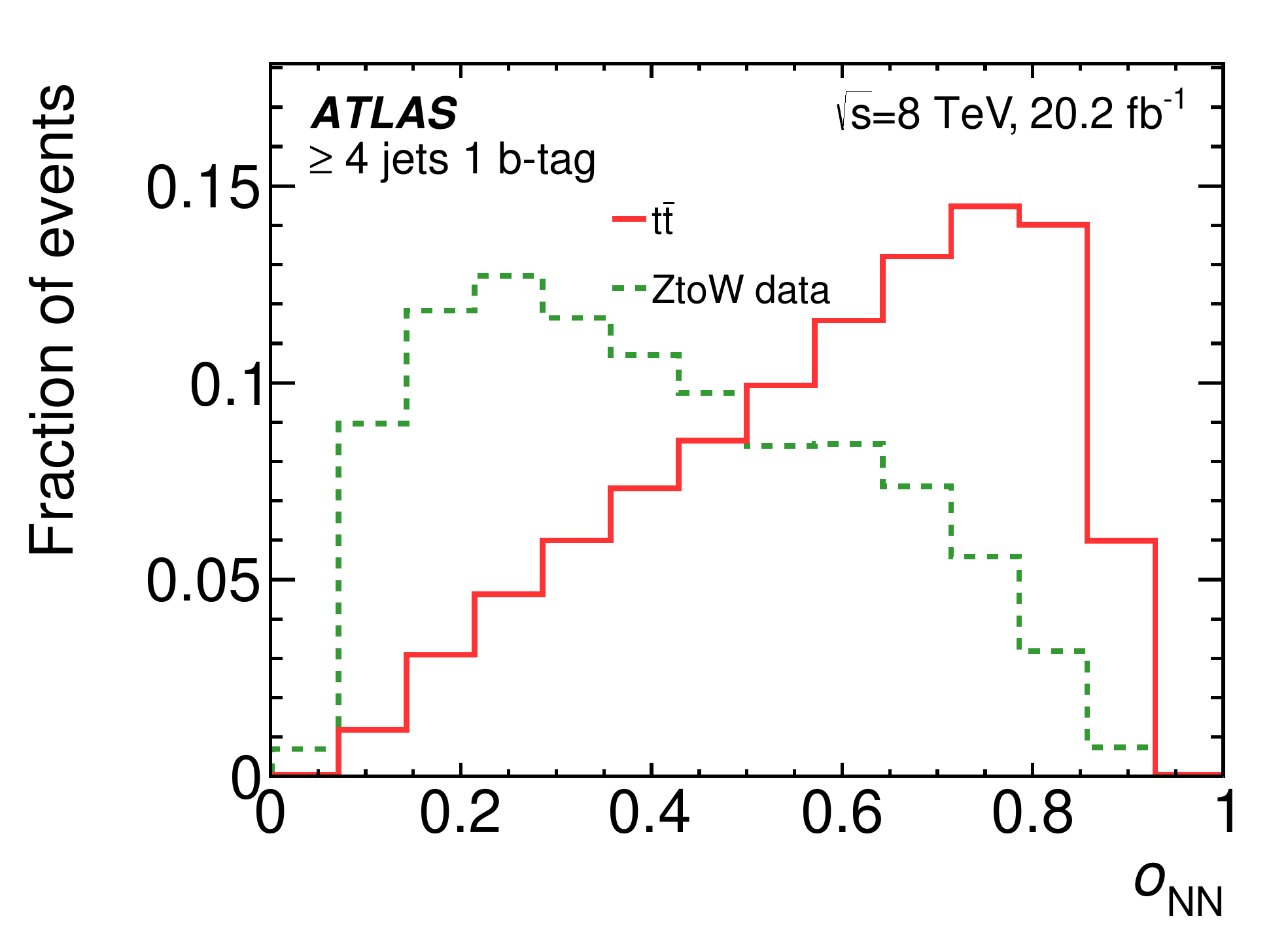}
	 \label{subfig:NN_shape_ch1} 
	}
    \subfigure[]{
    \includegraphics[width=0.45\textwidth]{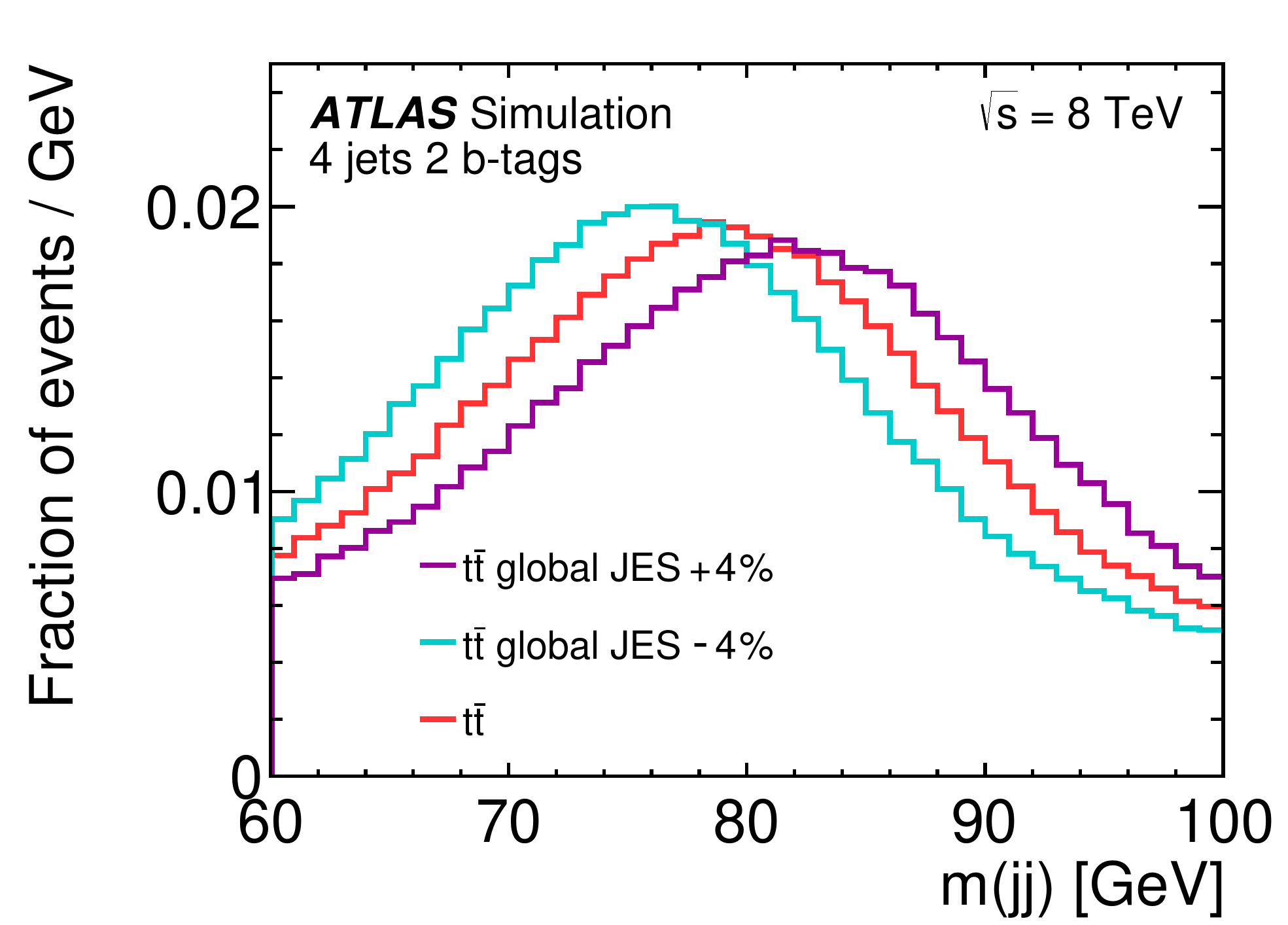} 
    \label{subfig:mjj_jes}
    }
	\caption{ 
	\subref{subfig:NN_shape_ch1} Probability densities of the neural-network
	discriminant $o_{\mathrm{NN}}$ for the simulated \ttbar signal process and the \Wjets background
	process derived from data using converted \Zjets events
	 for \SRoneNoSpace. \subref{subfig:mjj_jes} Probability densities for the
	 \ttbar signal process of the $m(jj)$ distribution for three different values of the JES, where events
    beyond the $x$-axis range are not shown and the range is restricted to show
    the peak~\cite{TOPQ-2016-08}.}
\end{figure}

For the determination of the \ttbar cross-section, a discriminant variable in
each signal region is defined. The number of \ttbar events is extracted using a
simultaneous fit of all three discriminant distributions to observed data. In
order to reduce systematic uncertainties due to the jet energy scale and
$b$-tagging efficiency, their effects on the signal and background distributions
are parametrised with nuisance parameters, which are included in the fit. The
ratio of single to double $b$-tagged events, i.e. the ratio of events in \SRone
and the sum of events in \SRtwo and \SRthr together is sensitive to the
$b$-tagging efficiency. In \SRone and \SRthrNoSpace, the output distribution of
an artificial neural network (NN)~\cite{Feindt:2006pm} is used.
Seven variables are chosen as input to the NN. These variables are based on
invariant masses between jets and leptons, event shape observables and
properties of the reconstructed top quarks. For the training of the NN, an equal
number of simulated \ttbar events and \ZtoW events is used. The discriminating
power of the NN between \ZtoW and \ttbar events can be seen in
Fig.~\ref{subfig:NN_shape_ch1} for \SRone. 
Since in \SRtwo the background
contribution is very small, a different distribution is used as the discriminant
in the final fit, which is the invariant mass of the two untagged jets $m(jj)$. The
$m(jj)$ is frequently utilised in several measurements of the top quark mass to
reduce the impact of the jet energy scale (JES) uncertainty~\cite{TOPQ-2013-02}.
The dependency of $m(jj)$ on the JES is shown in Fig.~\ref{subfig:mjj_jes} using
simulated \ttbar events with modified JES correction factors. Here the energy of
the jets are scaled by a constant scaling factor of $\pm \SI{4}{\percent}$.
\section{Result}
\begin{figure}[!ht]
\centering
    \subfigure[]{
    \includegraphics[width=0.3\textwidth]{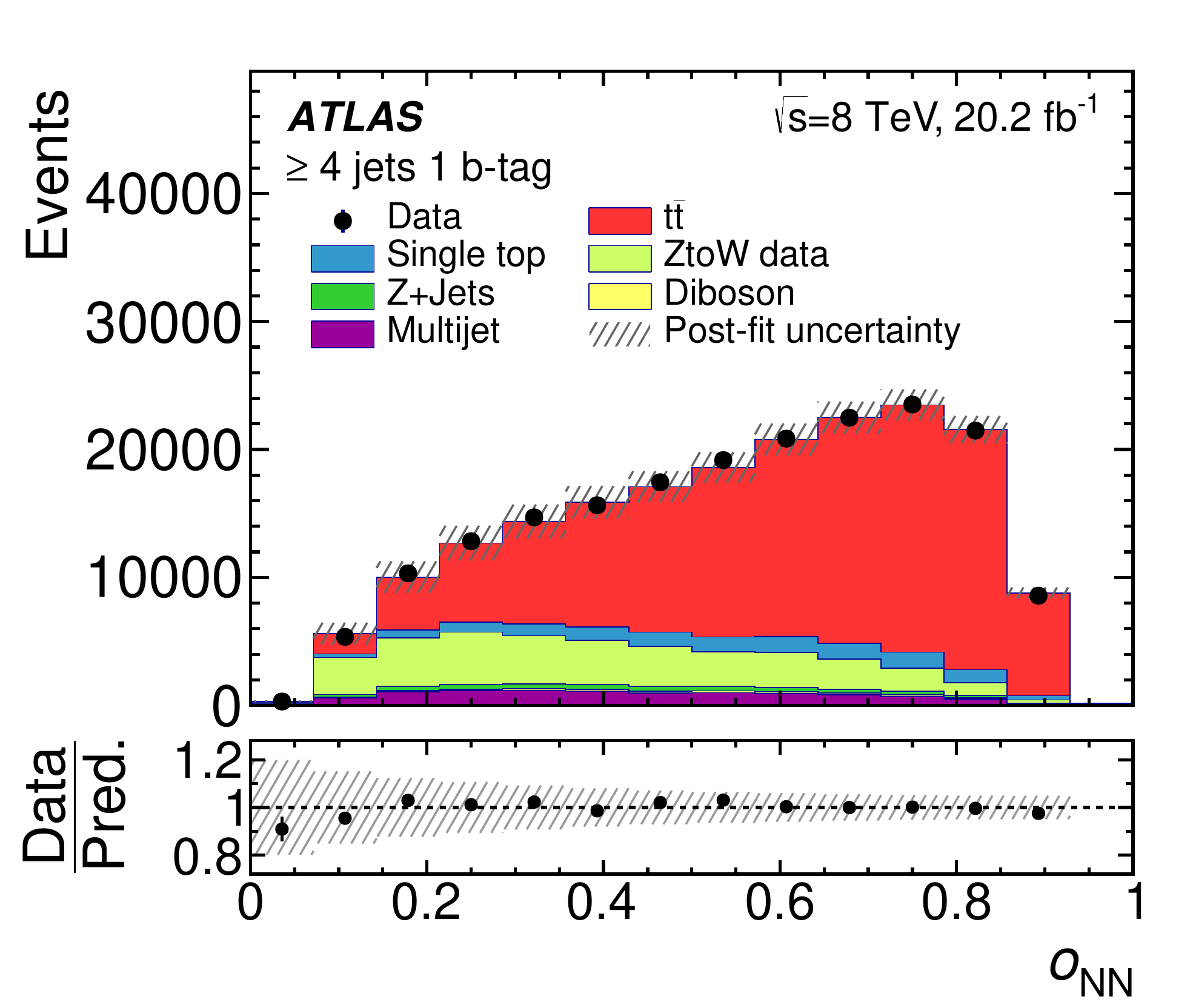}
    \label{subfig:postfit_nnnout_sr1}
    }
    \subfigure[]{
  \includegraphics[width=0.3\textwidth]{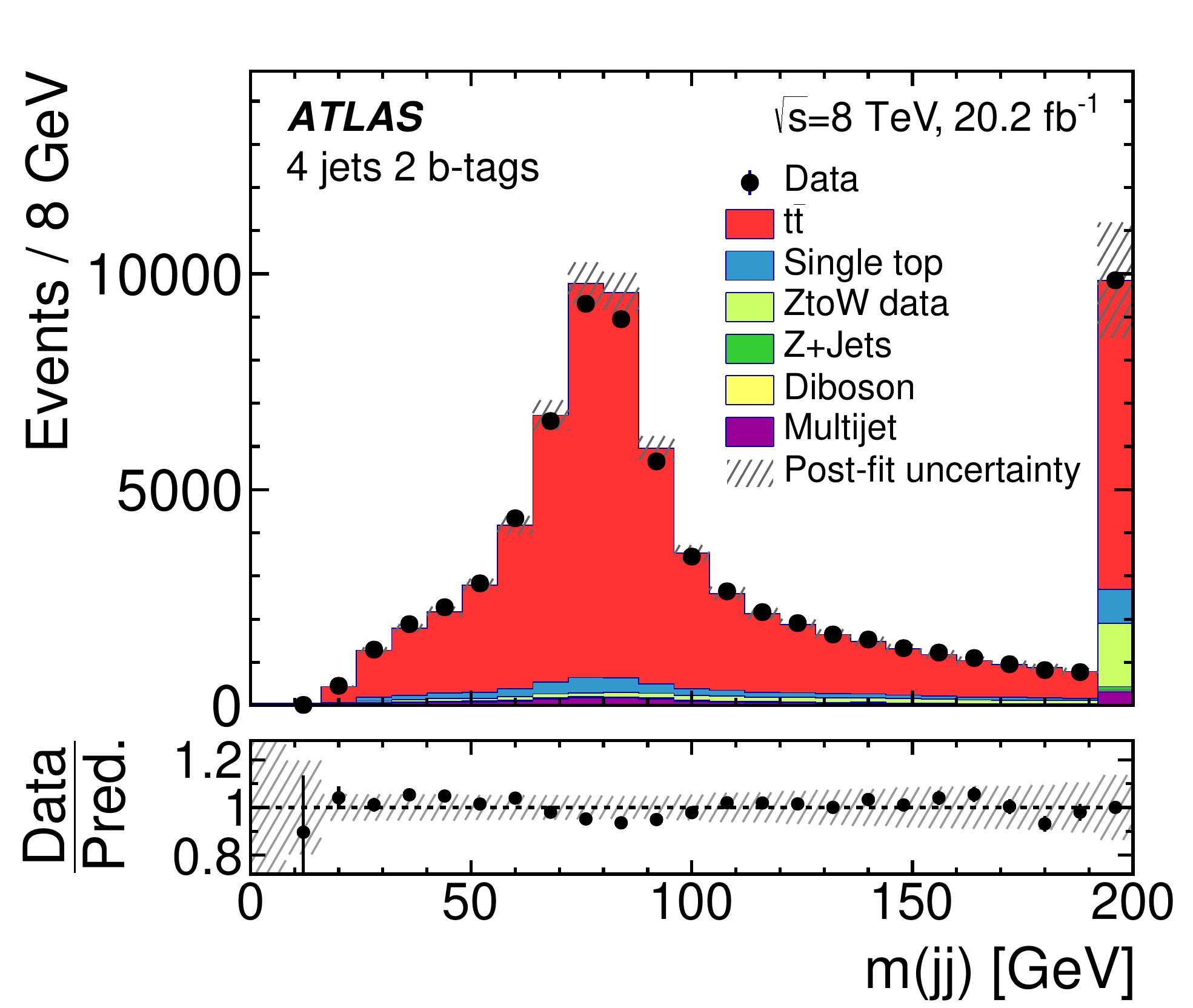}
  \label{subfig:postfit_mjj_sr2}
    }
    \subfigure[]{
  \includegraphics[width=0.3\textwidth]{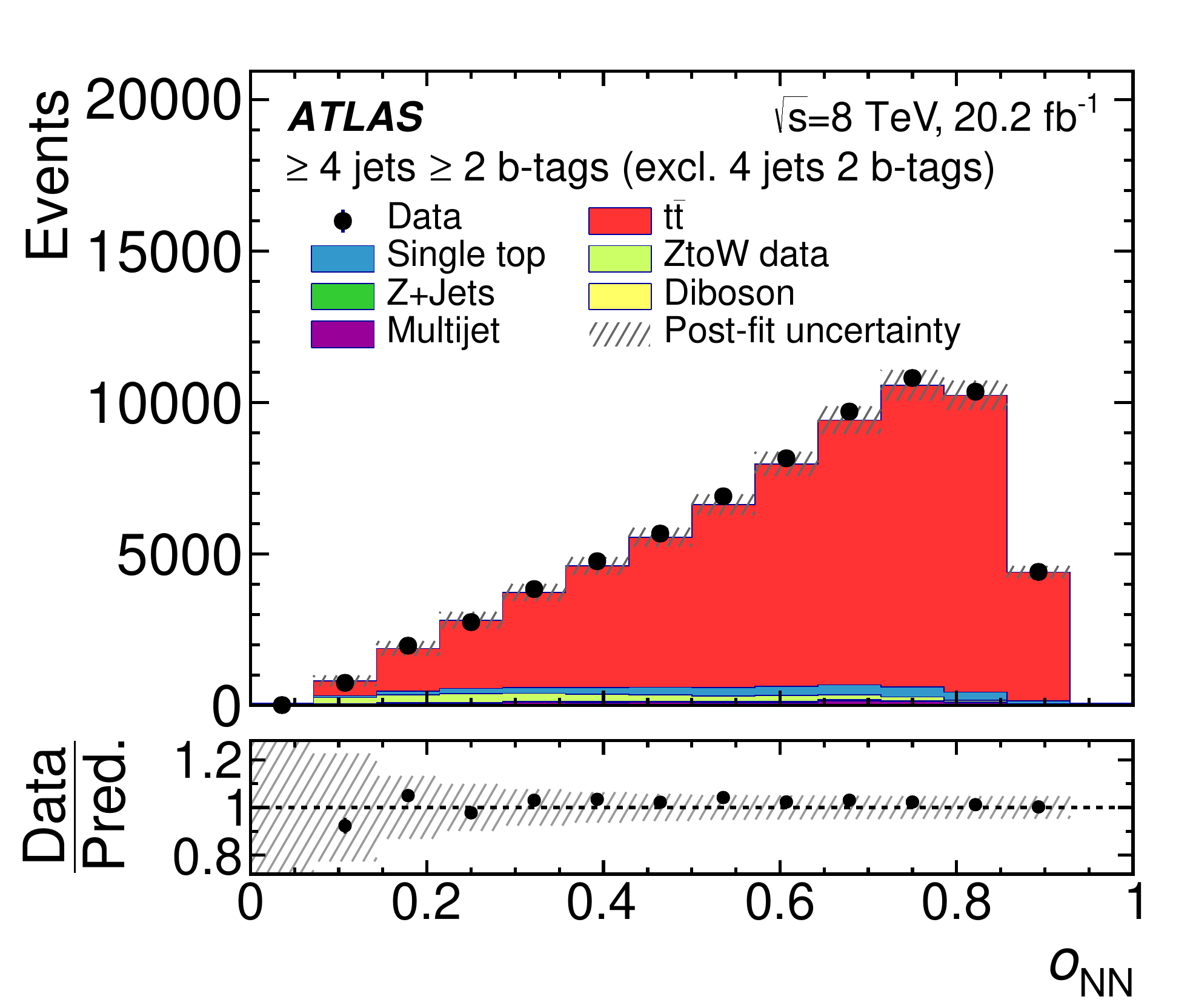}
  \label{subfig:postfit_nnnout_sr3}
    }
\caption{\label{fig:Fit}
Neural network discriminant $o_{\mathrm{NN}}$ and the $m(jj)$ distributions
normalised to the result of the maximum-likelihood fit for
\subref{subfig:postfit_nnnout_sr1} \SRoneNoSpace,
\subref{subfig:postfit_mjj_sr2} \SRtwoNoSpace, and
\subref{subfig:postfit_nnnout_sr3} \SRthrNoSpace.
     The hatched and gray error bands represent the post-fit uncertainty.
     The ratio of observed to predicted (Pred.) number of events in each bin is
     shown in the lower histogram. Events beyond the $x$-axis range are included
     in the last bin~\cite{TOPQ-2016-08}.}
\end{figure}

The distributions of signal and background processes scaled and morphed to the
fitted values are compared to the observed distributions of the NN discriminant
distribution in \SRone and \SRthr and the $m(jj)$ distribution in
\SRtwoNoSpace, shown in Fig.~\ref{fig:Fit}.
The total uncertainty in the inclusive \ttbar cross-section
is determined to be $\pm
\SI{5.7}{\%}$. The largest uncertainty is due to the uncertainty
in the PDF sets and the MC modelling of the signal process.
The uncertainties in the JES and the $b$-tagging efficiency have been
significantly reduced by including them as nuisance parameters together with the
choice of the signal regions and the discriminant distributions.

The inclusive \ttbar cross-sections is measured to be~\cite{TOPQ-2016-08}:
\begin{eqnarray*}
\sigma_{\ttbar} = 248.3 \pm 0.7 \, (\mathrm{stat.})  \pm 13.4 \,
(\mathrm{syst.}) \pm 4.7 \, (\mathrm{lumi.}) \ \mathrm{pb} 
\end{eqnarray*}
assuming a top-quark mass of $\mtop = \SI{172.5}{\GeV}$.

\bibliographystyle{atlasBibStyleWoTitle}
\bibliography{ATLAS.bib,ConfNotes.bib,document.bib}
 
\end{document}